\documentclass[twocolumn,prb,showpacs,multicol,amsmath,amssymb]{revtex4}
\usepackage[dvips]{graphicx}
\usepackage[dvipsnames]{xcolor}
\usepackage{graphicx}
\usepackage{dcolumn}
\usepackage{bm}
\usepackage{graphics}
\usepackage{epsfig,color}
\usepackage{color,soul}

\begin{document}
\title[J.\ Vahedi]{ Enhancing Thermoelectric Properties Through a Three-Terminal Benzene Molecule}
\author{Z.\ Sartipi$^{1}$ and J.\ Vahedi$^{*1,2}$}
\address{$^{1}$Department of Physics, Sari Branch, Islamic Azad University, Sari, Iran}
\address{$^{2}$ Laboratoire de Physique Theorique et Modelisation, CNRS UMR 8089, Universite de Cergy-Pontoise, 95302 Cergy-Pontoise Cedex, France.}
\date{\today}
\begin{abstract}
The thermoelectric transport through a benzene molecule with three metallic terminals is discussed. Using general local and non-local transport coefficients, we investigated different conductance and thermopower coefficients within the linear response regime. Based on the Onsager coefficients which depend on the number of terminal efficiencies, efficiency at maximum power is also studied. In the three-terminal set up with tuning temperature differences, a great enhancement of the figure of merit is observed. Results also show that the third terminal model can be useful to improve the efficiency at maximum output power compared  to the two-terminal model.
\end{abstract}
\pacs{85.65.+h, 85.80.Fi}
\maketitle
\section{Introduction}\label{sec1}
Thermoelectricity has recently received enormous attention due to the powerful ways of energy conversion. Enhancing the efficiency of thermoelectric materials, in the whole range spanning from macro- to nano-scales, is one of great importance for several different technological applications\cite{a2,a3,a4,a5}. The efficiency of a thermoelectric device consisting of two terminals is determined by its thermoelectric figure of merit ($ZT=\frac{GS^{2}}{K}T$), where $T$ is the temperature, $S$ is the Seebeck coefficient (thermopower), $G$ is the electrical conductance, and $K$ is the thermal conductance given by $K=K_p+K_e$ where $K_p$ ($K_e$) is the phononic (electronic) contribution to $K$. Clearly, $ZT$ can be increased by enhancing the power factor ($S^2GT$) or reducing the thermal conductivity and therefore a high-performance thermoelectric material should possess a large thermopower and electrical conductivity and simultaneously a low thermal conductivity. As these factors are correlated, increasing $ZT$ to values greater than unity is challenging. One promising approach has been to reduce the contribution of phonons by nanostructuring materials\cite{a7}.  
\par
While most of the researches have been conducted in two-terminal setups, transport in multi-terminal devices has begun to be investigated since these more complex designs may offer additional advantages\cite{a8,a9,a10,a11,a12,a13,a14,a15,a16,a17,a18}. An interesting perspective, for instance, is the possibility to exploit a third terminal in order to decouple the charge and energy flows and improve thermoelectric efficiency \cite{a20,a21,a19,a19b}.

\par
From an application point of view, the systems with high thermoelectric efficiency would be useful for waste energy harvesting\cite{a23,a24,a25}. Since the efficiency at maximum power monotonically depends on the thermoelectric figure of merit\cite{a26}, it is important to find heterostructures\cite{a27,a28,a29,a30} or bulk materials\cite{a31}, where the large thermopower guarantees high values of $ZT$. In this regard multi-terminal nanostructures offer enhanced flexibility and therefore it might be useful to improve efficiency\cite{a9,a32,a33,a34} especially under broken time-reversal symmetry\cite{a13,a20,a21}. In these structures, the non-equilibrium conditions are accompanied by significant non-local effects which require proper definitions of transport coefficients. Mazza et~al. derived an analytical expression for local and non-local transport coefficients and investigated how a third terminal could improve the performance of a quantum system\cite{a32}. In addition, Mechaleck et~al. propose an experimental protocol for determination of the transport coefficients in terms of the local and non-local conductances and thermopowers in three-terminal structures with a quantum dot\cite{a36}.  Erdman et~al., study the thermoelectric properties and heat-to-work conversion performance of an interacting, multi-level quantum dot (QD) weakly coupled to electronic reservoirs \cite{a37}.
\par
The nanostructuring of materials is also a crucial factor in the improvement of power factor because it can yield sharp features in the electronic density of states and the transmission coefficient which describe the propagation of electrons through a device. Even though these existed works, it is still desirable to explore new low-dimensional thermoelectric materials. In particular, the rapidly growing amount of research in atomic and molecular nanostructures should be notices. In this spirit, the use of single-molecule junctions has become an interesting topic in the field of molecular electronics. They are a powerful tool for detection of the intrinsic physical and chemical properties\cite{a38a,a38b,a38c,a38d,a38e,a39,a40,a41,a42}. The charge transport through the single benzene molecular junction has attracted a considerable attention and has become one of the most intensively studied nanoscale systems. Based on the benzene molecular junction, researchers able to manipulate single-molecule devices were in the Coulomb blockade regime \cite{a40}, observed the quantum interference effect \cite{a41} and studied the thermoelectric effects\cite{a42} in both experimental and theoretical methods. 
\par
We consider a benzene molecule for the central region since this configuration can highlight a series of complex interference effects due to multiple allowed pathways which can lead to Fano-like resonances in the transmission spectrum\cite{a44}. Note that quantum effects yielding Fano resonances in the transmission spectra were predicted to have an impact on the thermoelectric efficiency of single molecule devices\cite{a45}, nanoscale junctions\cite{a46} and quantum dot systems\cite{a47,a48,a49}. In addition to the dramatic effects of quantum interference on the performance of our system as a heat engine, our aim is to investigate how the efficiencies at maximum output power, output power, and figures of merit evolve when the system is driven from a two terminal to a three-terminal configuration.
\par
The rest of the paper is structured as follows: section-II provides theory and formalism. Results are presented in section-III and  section-IV gives a summary and conclusion.  
\begin{figure}
\includegraphics[width=9cm, height=6cm]{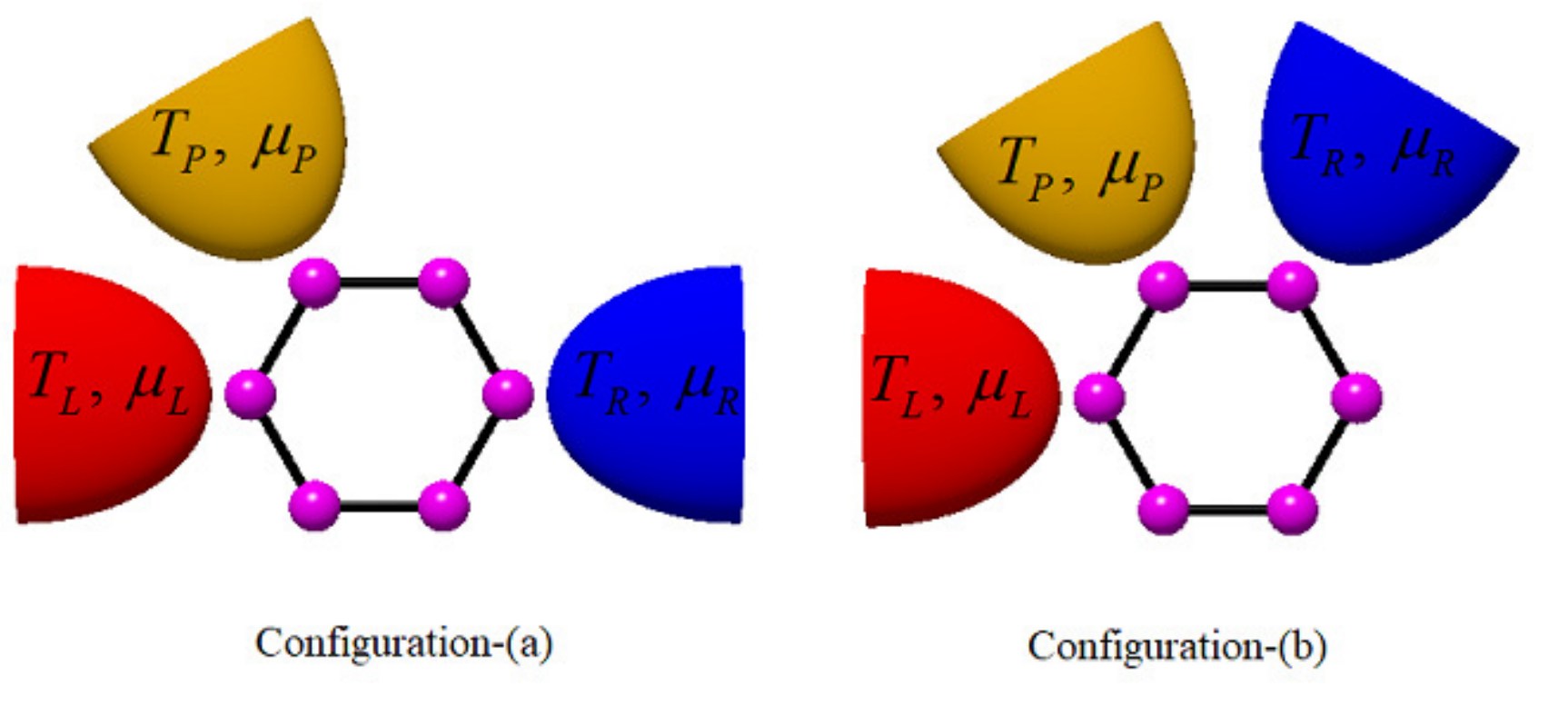}
\caption{A schematic view of the benzene molecule coupled to three nonmagnetic leads.}
\label{fig1}
\end{figure}

 \section{MODEL, THEORY And FORMALISM}\label{sec1}
Fig.\ref{fig1} shows a benzene molecule connected to three nonmagnetic leads, which can be described by the  Hamiltonian as follows $$\mathcal{H}=\mathcal{H}_{leads}+\mathcal{H}_m+\mathcal{H}_{tun}$$  where $\mathcal{H}_{leads}=\sum_{\alpha k}\varepsilon_{\alpha k}c_{\alpha k}^{\dagger}c_{\alpha k}$ is the Hamiltonian of three ($\alpha=R, P, L$) leads connected to the benzene molecule. $c_{\alpha k}^{\dagger}$ creates an electron in the $ \alpha= L,P,R$ lead with  energy $\varepsilon_{k}$. The Hamiltonian of the benzene molecule is $\mathcal{H}_m=\sum_i\varepsilon_{i}d_i^{\dagger}d_i+t\sum_{ij}(d_i^{\dagger}d_j +h.c)$, where $d_i^{\dagger}(d_i)$ is the creation (annihilation) operator of an electron at site $i$ in the benzene molecule, $ \varepsilon_i$ and $t$ describe the on-site energy and the nearest-neighbour hopping integral in the benzene molecule, respectively.  The last term is the tunneling between leads and molecule $\mathcal{H}_{tun}=\sum_{\alpha ki}\gamma_{\alpha ki} (c_{\alpha k}^{\dagger}d_{i}+h.c)$, where $\gamma_{\alpha ki}$ represents the coupling strength between the molecule and leads.   For a scattering region including a benzene molecule in contact with three leads, we can express the transmission probability as ${{\cal T}(E)}=Tr(\Gamma_{i}G^{r}\Gamma_{j}G^{a})$, where  $\Gamma_{\alpha}=i(\Sigma_{\alpha}-\Sigma_{\alpha}^{\dagger})$, $(\alpha=L,P,R)$ is the broadening matrix due to the coupling to leads, where $\Sigma_{\alpha}$ being the self-energy of lead $i$. $ G^{a} $ and $ G^{r} $ are advanced and retarded green function of the molecule, respectively. Note that in the wide band limit approximation, we set $\Sigma_{\alpha}=-i\frac{\gamma_{\alpha}}{2}$, where $ \gamma_{\alpha} $ does not depend on energy. This choice yields the identification $\Gamma_{\alpha}=\gamma_{\alpha} $.
 \par
 \subsection{Non-equilibrium thermodynamics}
The system under consideration (see Fig.\ref{fig1}) is characterized by three energy and  particle currents $J^U_{\alpha}$ and $J^N_{\alpha}$ $({\alpha}=L, R, P)$, respectively, which flow from the corresponding terminals, in accordance with the constraint $\sum_{\alpha}J_{\alpha}^{U(N)}=0$. Note that positive values correspond to flows from the terminals to the system. We will set electrode $R$ as a reference $(T_R,\mu_R)\equiv(T,\mu)$. To guarantee the linear response regime is considered, we set $(T_{\alpha},\mu_{\alpha})\equiv(T+\delta T_{\alpha},\mu+\delta\mu_{\alpha})$ with $\left|\delta\mu_{\alpha}\right|/k_BT\ll1$ and $|\delta T_{\alpha}|/T\ll1$ for ${\alpha}=L, P$, where $k_B$ is the Boltzmann constant. 
Two thermodynamic forces $X_{L,P}^{\mu}$ and $X_{L,P}^T$  and their fluxes $J^N_{L,P}$ and $J^Q_{L,P}$ response  can describe the non-equilibrium thermodynamics processes \cite{a51,a52}. During the processes, total charge $x$ transfer from one terminal to another one is done by work $W=-Fx$ against an external thermodynamic force $F$.  The corresponding force is $X_{L,P}^{\mu}=F/T$ which $T$ being the temperature of the system and $F=\delta\mu_{L,P}/e$, where $e$ is the electron charge.  The thermodynamic flux is defined as $J^N_{L,P}=dx/dt$.  We are interested in the heat to work conversion, that is, the work is performed by converting  part of the heat $Q$ which flows from the hot terminals.  Regarding the linear regime, the temperature difference $\delta T_{L,P}=T_{L,P}-T$ is small compared to $T_L\approx~T_P\approx~T_R$, thus the thermodynamic force is $X^T_{L,P}=\delta T_{L,P}/T^2$, and the heat current is $J^Q_{L,P}=dQ/dt$. 
 \par
For a three-terminal setup, the elements of the Onsager matrix is related to eight quantities (details are presented in Appendix \ref{apeA}). The matrix elements can be categorized in two local and non-local thermopower $S_{\alpha\beta}$, the electrical $G_{\alpha\beta}$ and thermal $K_{\alpha\beta}$ conductances. The non-local $ (\alpha\neq\beta) $ coefficients describe how bias driven between two terminals can influence transport in another terminal . We follow Ref.[~\onlinecite{a32}] terminology, so let's define the transport coefficients for a three-terminal system as follows 
\begin{eqnarray}
S_{LL}&=&\frac{1}{eT}\frac{{{\cal L}^{(2)}_{13;32}}}{{{\cal L}^{(2)}_{13;31}}},\quad\quad S_{PP}=\frac{1}{eT}\frac{{{\cal L}^{(2)}_{14;31}}}{{{\cal L}^{(2)}_{13;31}}}\nonumber\\
S_{LP}&=&\frac{1}{eT}\frac{{{\cal L}^{(2)}_{13;34}}}{{{\cal L}^{(2)}_{13;31}}},\quad\quad S_{PL}=\frac{1}{eT}\frac{{{\cal} {{\cal L}^{(2)}_{13;21}}}}{{{\cal L}^{(2)}_{13;31}}}.
\label{eq1}
\end{eqnarray}
\begin{eqnarray}
G_{LL}&=&\frac{e^2 {{\cal L}_{11}}}{T},\quad\quad G_{PP}=\frac{e^2 {{\cal L}_{33}}}{T}\nonumber\\
G_{LP}&=&\frac{e^2{{\cal L}_{13}}}{T},\quad\quad G_{PL}=\frac{e^2{{\cal L}_{13}}}{T}.
\label{eq2}
\end{eqnarray}
\begin{eqnarray}
K_{LL}&=&\frac{1}{T^2}\frac{{{\cal L}_{12}}L^{(2)}_{12;32}-{{\cal L}_{12}}L^{(2)}_{13;32}-{{\cal L}_{11}}{{\cal L}^{(2)}_{23;23}}}{L^{(2)}_{13;31}}\nonumber\\
K_{PP}&=&\frac{1}{T^2}\frac{{{\cal L}_{14}}{{\cal L}^{(2)}_{13;43}}-{{\cal L}_{13}}{{\cal L}^{(2)}_{14;43}}-{{\cal L}_{11}}{{\cal L}^{(2)}_{34;34}}}{{{\cal L}^{(2)}_{13;31}}}\nonumber\\
K_{LP}&=&\frac{1}{T^2}\frac{{{\cal L}_{24}}{{\cal L}^{(2)}_{13;31}}- {{\cal L}_{14}} {{\cal L}^{(2)}_{13;23}}-{{\cal L}_{34}}{{\cal L}^{(2)}_{13;12}}  }{{{\cal L}^{(2)}_{13;31}}}\nonumber\\
K_{PL}&=&K_{LP}
\label{eq3}
\end{eqnarray}
where ${\cal L}^{(2)}_{ij,kl}$=${\cal L}_{ik} {\cal L}_{lj}- {\cal L}_{il}{\cal L}_{kj}$. It is worth also that  by disconnecting terminal $P$ from the rest,  two-terminal well known formula for Seebeck coefficient, electrical and thermal conductance can be recovered as $ S_{LL} $=$\frac{1}{eT}\frac{{\cal L}_{12}}{{\cal L}_{11}}$, $ G_{LL}$= $\frac{e^{2}}{T}{{\cal L}_{11}}$ and $ K_{LL} $=$\frac{1}{T^{2}}\frac{{\cal L}^{(2)}_{12;12}}{{\cal L}_{11}}$, respectively. The Onsager coefficients ${\cal L}_{ij}$ are obtained by the linear response expansion of the currents $J_i$ ( see Appendix \ref{apeC}).
\subsection{Efficiency at maximum output power}
We aim to create a heat engine, so we should  consider its efficiency. But we just interested  in the efficiency at maximum output power $\eta(\textbf{P}_{max})$. Indeed, we look for the value of the efficiency at chemical potential which optimizes the output power generated by the heat engine.  Thus we first need to define the output power $ \textbf{P}$=$\frac{dW}{dt}$ where $W$ denotes the work performed by the system against an external force $F$. This can be recast base on generalized forces  and fluxes as  $ \textbf{P}=-T(J_{L}^{N}X_{L}^{\mu}+J_{P}^{N}X_P^{\mu})$
\par
Having the output power the  the steady-state efficiency of the heat engine can be defined as the output power \textbf{P}, divided by the sum of the heat currents absorbed by the engine $\eta=\frac{\textbf{P}}{\sum_{\alpha}J_{\alpha}^Q}$, with constrain of positive heat currents in  the $\sum_{\alpha} $ in the denominator .  Note that the  signs of the heat currents depend on the details of the system, thus the efficiency depends on which heat currents are positive. In the three-terminal model considered in this work ( see Fig.\ref{fig1}), we focus on the situation where $ J_R^Q$ is negative. So the efficiency can be defined as $ \eta_{LP}=\frac{\textbf{P}}{J_L^Q+J_P^Q} $ when both $ J_L^Q$ and $J_P^Q$ are positive and given as $\eta_{L(P)}=\frac{\textbf{P}}{J_{L(P)}^Q} $ when either $J_L^Q$ or $J_P^Q$ is positive\cite{a32}. For the sake of simplicity and without loss of generality, we set $T_L>T_P>T_R$ , thus $ \delta{T}_L>\delta{T}_P>0 $. 

\par
The electrochemical forces $ X_{L,P}^{\mu} $ that maximize output power at given temperature forces $ X_{L,P}^{T} $ can be written as
\begin{eqnarray}
X_L^{\mu}&=&\frac{-eT}{2}\left( S_{LP}X_P^T+S_{LL}X_L^T\right) \nonumber\\
X_P^{\mu}&=&\frac{-eT}{2}\left( S_{PL}X_L^T+S_{PP}X_P^T\right) 
\label{eq4}
\end{eqnarray}
inserting these expressions into the output power yields 
\begin{eqnarray}
\textbf{P}_{max}&=&\frac{T^4}{4}\left({\cal X}^{\dagger}~{\cal M}~{\cal X}\right)
\label{eq5}
\end{eqnarray}
where $ {\cal X}^{\dagger}=\begin{bmatrix}
{ X_L^T} & {X_P^T} 
\end{bmatrix}$ 
 and ${\cal M}= \begin{bmatrix}
{\cal M}_{11} & {\cal M}_{12} \\
{\cal M}_{21} & {\cal M}_{22}
\end{bmatrix}$ is a positive semi-definite matrix, whose elements are defined as follows, 
\begin{eqnarray}
 {\cal M}_{12}&=& {\cal M}_{21}=G_{LP}S_{LP}S_{PL}+G_{LP}S_{LL}S_{PP}\nonumber\\
 &+&G_{PP}S_{PL}S_{PP} +G_{LL}S_{LL}S_{LP}\nonumber\\
{\cal M}_{11} &=&G_{LL}S_{LP}^2+2~G_{LP}S_{LP}S_{PP}+G_{PP}S_{PP}^2\nonumber\\
{\cal M}_{22}&=&G_{LL}S_{LL}^2+2~G_{LP}S_{PL}S_{LL}+G_{PP}S_{PL}^2
\label{eq6}
\end{eqnarray}
 inserting ${\cal M}$ into $\textbf{P}_{max}$ gives 
\begin{eqnarray}
\textbf{P}_{max}&=&\left(\frac{T^{4}}{4} \right) {\cal N}^{2} \left( \lambda_1\cos^{2}\theta+ \lambda_{2}\sin^{2}\theta\right)     
\label{eq7}
\end{eqnarray}
where $ {\cal N}=\|(X_{L}^{T}~X_{P}^{T} ) \|$ is the system temperature and the angle $ \theta $ determines the rotation in the $(X_L^T~X_P^T )$ plane that defines the eigenvectors of ${\cal M}$. 
Here the parameter $ \rho= \left( \lambda_1\cos^{2}\theta+ \lambda_2\sin^{2}\theta\right) $ has the meaning of power factor for   three-terminal system which relates the maximum output power to the temperature differences ( for more details we refer interested readers to Ref.~[\onlinecite{a32}]). 
\par 
Now using Eq.(\ref{eq7}), one can obtain the efficiency at maximum output power $ \eta{(\textbf{P}_{max})}$. Analog to the equation of efficiency at maximum power for two-terminal cases, for three the terminal configuration,  the efficiency at maximum output power can also be expressed in terms of the Carnot efficiencies \cite{a32} as follows
\begin{widetext}
\begin{eqnarray}
{\eta_{L}}{(\textbf{P}_{max})}&=&\dfrac{{\eta_{c}}_{L}}{2}\dfrac{Z_{L}^{{\cal M}_{11}}T+2{\cal K}Z_{L}^{{\cal M}_{12}}T+{\cal K}^{2}Z_{L}^{{\cal M}_{22}}T}{2\frac{{\cal F}_1}{{\cal F}}+4{\cal K}{\cal F}_1+2{\cal K}^{2}+Z_{L}^{{\cal M}_{11}}T+2{\cal K}Z_{L}^{{\cal M}_{12}}T+{\cal K}^{2}Z_{L}^{{\cal M}_{22}}T}=\dfrac{{\eta_{c}}_{L}}{2}\dfrac{{\cal Z}_{L}T}{{\cal W}_L+{\cal Z}_L T}\nonumber\\
{\eta_{P}}{(\textbf{P}_{max})}&=&\dfrac{{\eta_{c}}_{P}}{2}\dfrac{Z_{P}^{{\cal M}_{11}}T+2{\cal K}Z_{P}^{{\cal M}_{12}}T+{\cal K}^2Z_{P}^{{\cal M}_{22}}T}{2{\cal K}^2\frac{{\cal F}}{{\cal F}_1}+4{\cal K}{\cal F}+2+Z_{P}^{{\cal M}_{11}}T+2{\cal K}Z_{P}^{{\cal M}_{12}}T+{\cal K}^2Z_{P}^{{\cal M}_{22}}T}=\dfrac{{\eta_{c}}_{P}}{2}\dfrac{{\cal Z}_{P}T}{{\cal W}_P+{\cal Z}_{P}T}\nonumber\\
{\eta_{LP}}{(\textbf{P}_{max})}&=&\dfrac{{\eta_{c}}_{LP}}{2}\dfrac{Z_{LP}^{{\cal M}_{11}}T+2{\cal K}Z_{LP}^{{\cal M}_{12}}T+{\cal K}^2Z_{LP}^{{\cal M}_{22}}T+{\cal O}(\delta T)}{2{\cal F}^{-1}+4{\cal K}+2{\cal F}^{2}{\cal K}_1^{-1}+Z_{LP}^{{\cal M}_{11}}T+2{\cal K} Z_{P}^{{\cal M}_{12}}T+{\cal K}^2Z_{P}^{{\cal M}_{22}}T+{\cal O}(\delta T)}\simeq\dfrac{{\eta_{c}}_{LP}}{2}\dfrac{{\cal Z}_{LP}T}{{\cal W}_{LP}+{\cal Z}_{LP}T}
\label{eq8}
\end{eqnarray}
\end{widetext}
where the constants,  ${\cal W}_L$=$2\frac{{\cal F}_1}{{\cal F}}+4{\cal K}{\cal K}_1 +2{\cal K}^2$,
${\cal W}_P$=$2{\cal K}^2\frac{{\cal F}}{{\cal F}_1}+4{\cal K}{\cal F}+2$ and 
${\cal W}_{LP}$=${\cal F}_1^{-1}+{\cal K}^2{\cal F}^{-1}+2{\cal K}$. ${\eta_c}_{LP}$, ${\eta_c}_{P}$ and ${\eta_c}_{L}$ are corresponding Carnot efficiencies.  One has  defined the parameters $ {\cal K}$=$\frac{X_L^T}{X_P^T}$, $ {\cal F}$=$\frac{K_{LP}}{K_{PP}}$ and $ {\cal F}_1$=$\frac{K_{LP}}{K_{LL}}$ . And the combination of the figure of merit
\begin{eqnarray}
 {\cal Z}_{\alpha}T&=&\left(Z_{\alpha}^{{\cal M}_{11}}+2{\cal K} Z_{\alpha}^{{\cal M}_{12}}+{\cal K}^2Z_{\alpha}^{{\cal M}_{22}}\right) T, \quad  ({\alpha}=L,P)\nonumber\\
{\cal Z}_{LP}T&=&\left(Z_{LP}^{{\cal M}_{11}}+2{\cal K}Z_{LP}^{{\cal M}_{12}} +{\cal K}^2Z_{LP}^{{\cal M}_{22}}\right)T
\label{eq9}
\end{eqnarray}
 where $ Z^j_{\alpha}T ~(j$=${\cal M}_{11},{\cal M}_{12},{\cal M}_{22})$ are generalized figures of merit defined as $ Z_L^jT=\frac{jT}{K_{LL}}T$, $ Z_P^jT=\frac{jT}{K_{PP}}$ and $Z_{LP}^jT=\frac{jT}{K_{LP}}$ . For detailed calculations, we refer readers to Ref.~[\onlinecite{a32}].

 \section{RESULTS AND DISCUSSION}\label{sec1}
 In this section, we present the numerical results. The parameters used are as follows: nearest neighbor hopping energy in the Benzene molecule is set as $t=1.5 eV$, and the coupling are considered constant in the wideband approximation.  For temperature setup we use a parameter $M$ which is defined by $\delta T_P=M\delta T_L$. Room temperature $T=300$ is chosen as the reference temperature. 
 \par
\begin{figure}[t]
 \includegraphics[width=9cm, height=5cm]{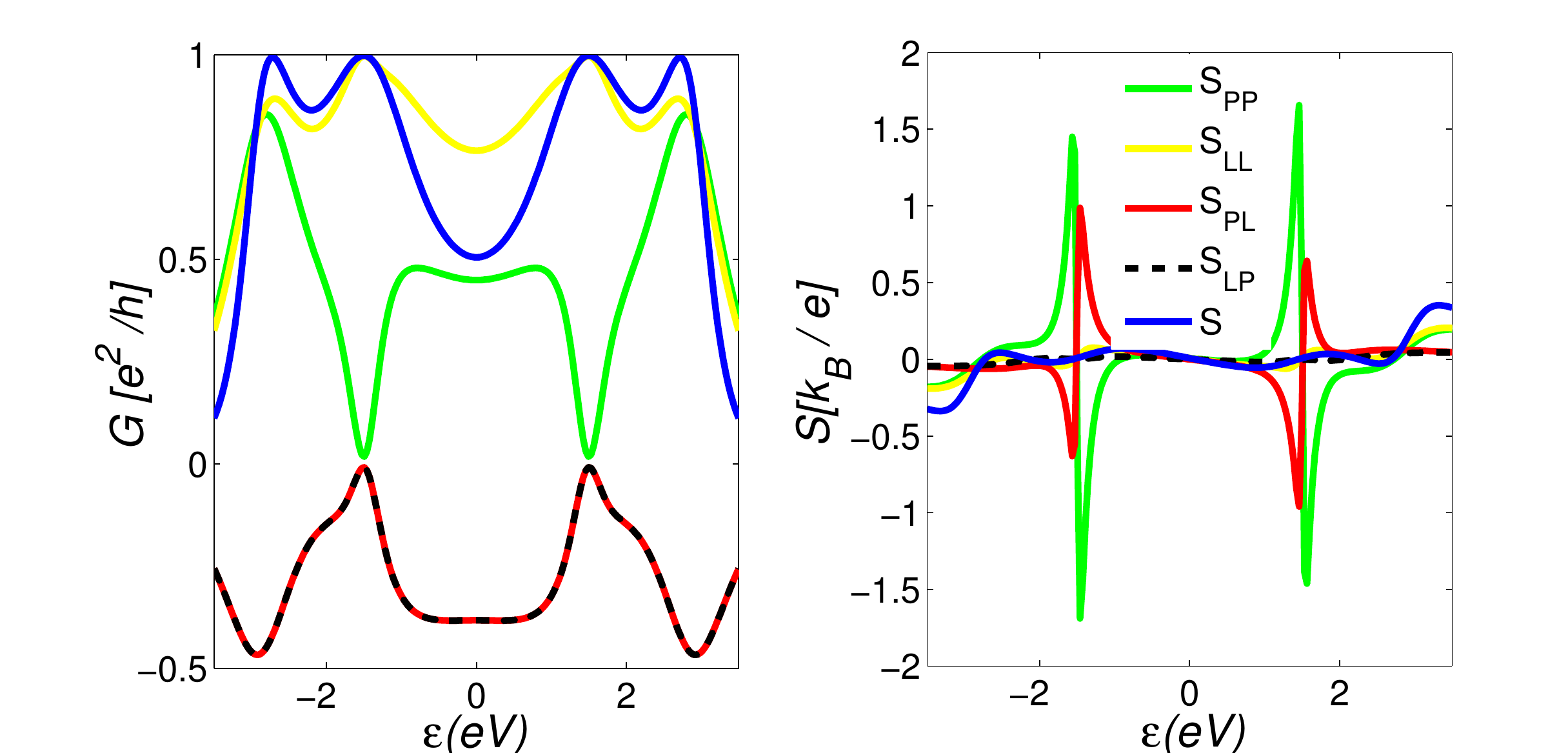}
 \caption{(Color online) Different thermopowers $S_{\alpha\beta}$, ($\alpha=L,P$), with local $ S_{LL}$ (pink line), $S_{PP}$ (green line) and non-local $S_{LP}$ (dashed black line),  $S_{PL}$ (red line) are plotted in left panel, and corresponding electrical conductances $G_{\alpha\beta}$ are plotted in right panel. The blue line shows the results of the two-terminal case. The temperature difference is fixed by $M=0.1$ and couplings are considered in a symmetric way as $\gamma_{P} =\gamma_{L}=\gamma_{R} =2.5 eV$.} 
\label{fig2}
\end{figure}

 Results of conductance and Seebeck coefficients as a function of incoming electron energy are plotted in Fig.~\ref{fig2}. For comparison purpose, the two terminal model also considered (shown in blue line). Two cases of local and non-local coefficients are labeled with $A_{\alpha\alpha}$ and $A_{\alpha\beta}$, respectively, where $A=G,S$ and $\alpha,\beta=L,R,P$. We should also note that the results depicted in Fig.~\ref{fig2} are based on the configuration-$(a)$ (see Fig.~\ref{fig1}) and we fix $M=0.1$. It can be seen that for the two-terminal case, the electrical conductance (blue line) shows four resonance peaks located at $\varepsilon=\pm 1.5 eV$ and $\varepsilon=\pm 3 eV$. Presence of a third terminal gives rise two types of local and non-local coefficients.  Except for $G_{LL}$, the rest conductances show two Fano-like resonances. One can explain this feature based on the quantum interferences effect caused by different pathways of electrons due to the coupling of the third terminal to the Benzene molecule.
\par
 It is also believed that a sharp change in the conductance characteristic would lead to a sharp enhacenment of thermopower (Seebeck). For two terminal case at low temperature this can be easily explained by Mott's formula $S=-\frac{\pi^2}{3}\frac{k_B^2T}{e}\frac{\sigma^{\prime}(\mu)}{\sigma}$. It can be seen that for the three terminal set up a great enhancement takes place for $S_{PP}$ and $S_{PL}$.
 \begin{figure}[t]
 \includegraphics[width=9cm, height=5cm]{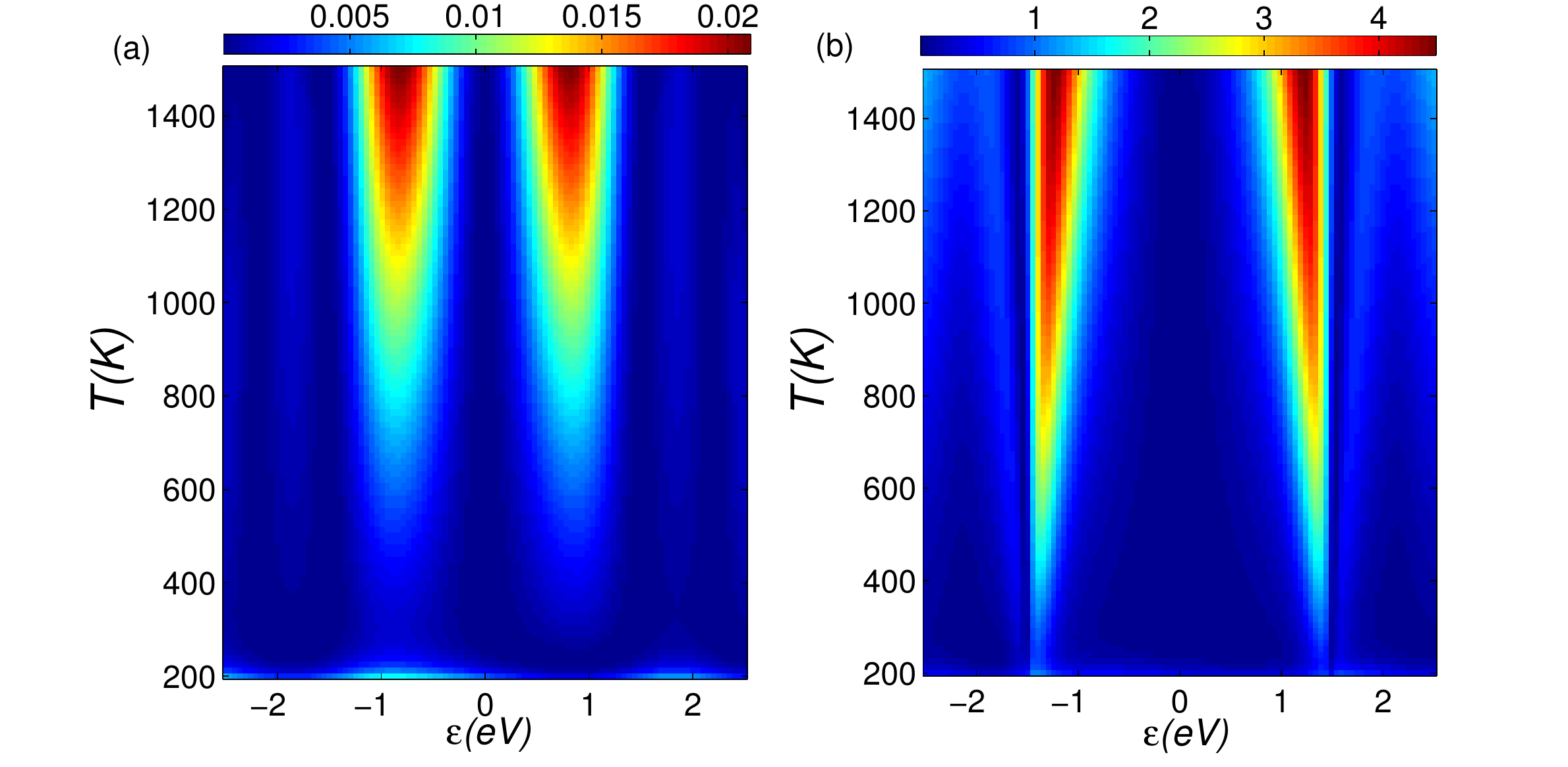}
 \caption{(Color online) Density plots of figures of merit $ZT$ for a two-terminal configuration (left panel) and ${\cal Z}_LT$ for configuration-$(a)$ (right panel) as a function of level position and temperature. Parameters are the same as Fig.~\ref{fig2}}
\label{fig3}
\end{figure}
\begin{figure}[t]
 \includegraphics[width=9cm, height=5cm]{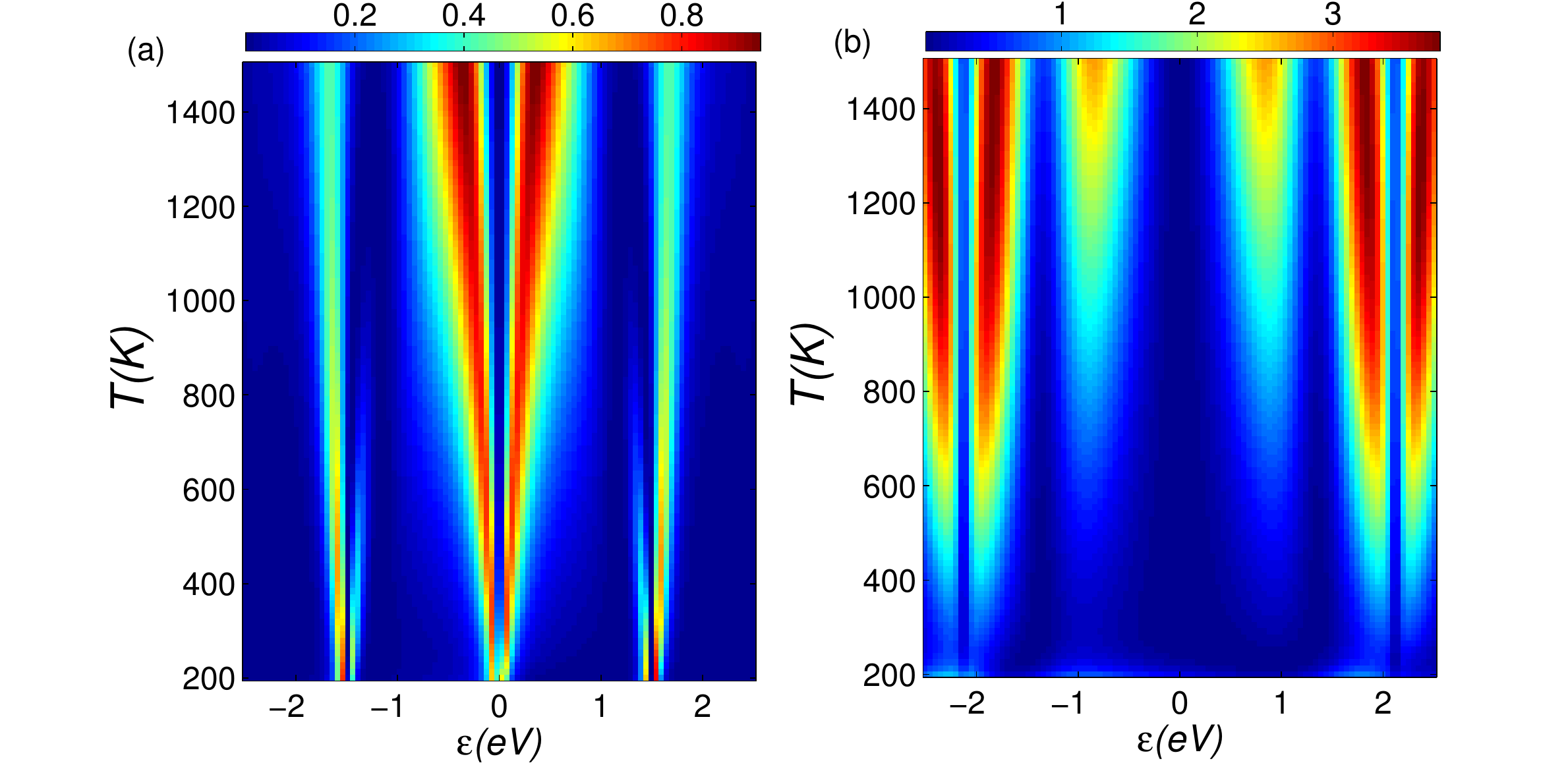}
 \caption{(Color online) Same as Fig.~\ref{fig3}, but for configuration-$(b)$} 
  \label{fig4}
\end{figure}
\par
Now, we consider temperature influence on the figure of merit particularly focusing on the efficiency at maximum power. For a two-terminal configuration, we use the well known efficiency at maximum power formula as $\eta^{\Vert} (P_{max})=\frac{\eta_{C}^{\Vert}}{2}\frac{ZT}{2+ZT}$, where $ZT=\frac{GS^2T}{K}$ is the  two-terminal figure of merit. For the three terminal case, we use Eq.(\ref{eq8}). Indeed, in a numerical approach, after fixing parameters, we monitor the heat currents $J^Q_{\alpha}$ with $(\alpha=L, R, P)$ then regarding the positivity of heat current, the corresponding efficiency and finally, the figure of merit is calculated. We use the same parameters as Fig.~\ref{fig2}. Left and right panels of Fig.~\ref{fig3} show the density plot of the figure of merit as a function of on-site energy $\varepsilon$ and temperature for a two-terminal configuration and configuration-$(a)$. One can see that the value of ${\cal Z}_LT$ for the three terminal case (right panel) is significantly enhanced in the vicinity of $\varepsilon=\pm1$ and reaches a value $\simeq 4.5$. 
\begin{figure}[t]
\includegraphics[width=9cm, height=5cm]{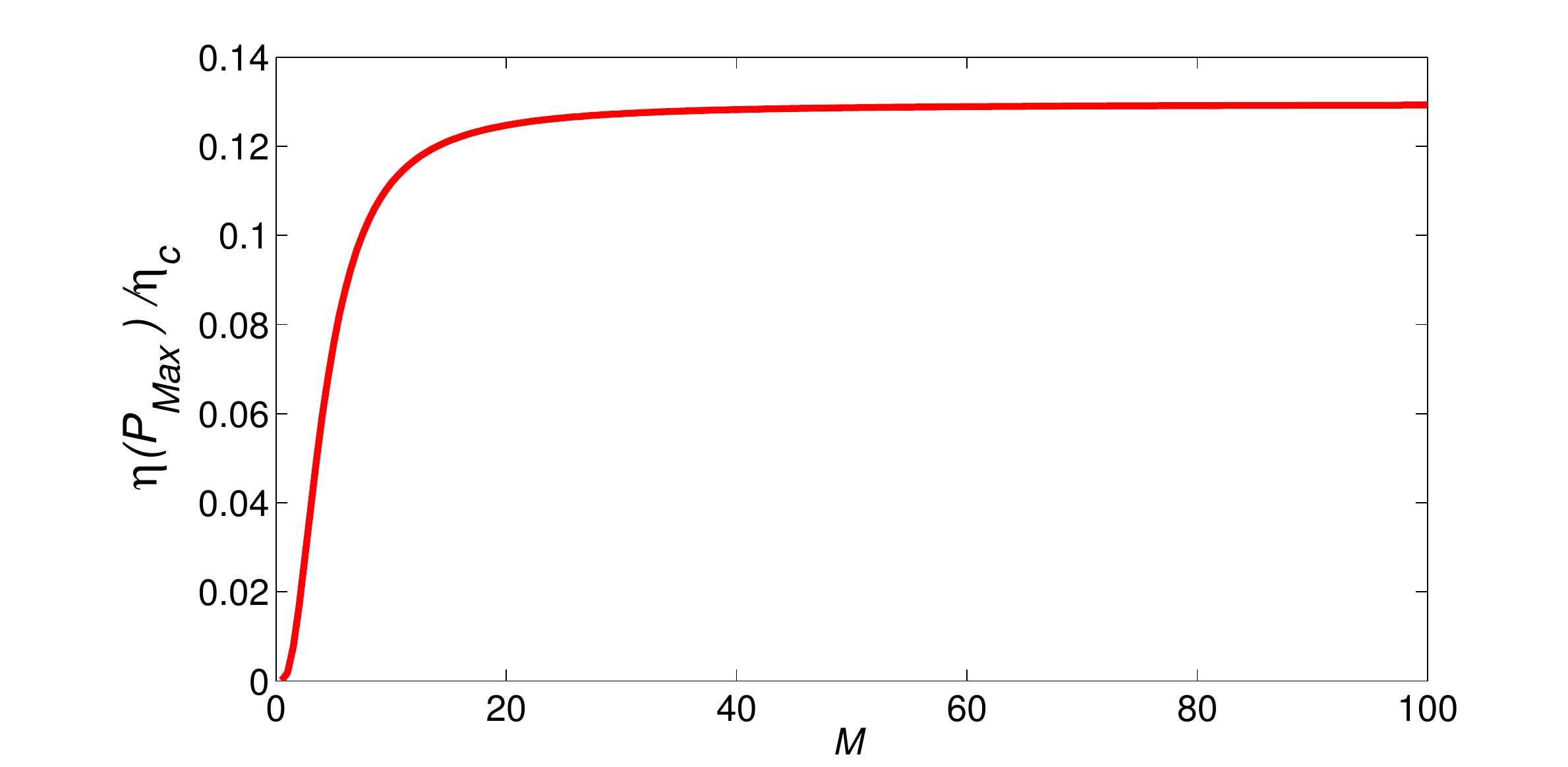}
\caption{(Color online) Efficiency at maximum output power normalized over Carnot efficiency as a function of $M$ for configuration-$(a)$. For $ 0\leq M<0.57 $ the system absorbs heat only from reservoir $L$ ($J^Q_L>0,~J^Q_{R,P}<0$) ; for $ 0.57 \leq M<18.5 $ the system absorbs heat from reservoirs $L$ and $P$  ($J^Q_R<0,~J^Q_{L,P}>0$); and finally, for $18.5\leq M\leq100$ the system absorbs heat only from reservoir $P$ ($J^Q_P>0~J^Q_{R,L}<0$).}
\label{fig5}
\end{figure}
\par
We also consider the thermoelectric properties of configuration-$(b)$  (see Fig.~\ref{fig1}). Among many choices, as the preceding paragraph, we focus on the figure of merit.  Fig.~\ref{fig4}, by tuning $M=0.1$ while the temperature difference between reservoirs $L$ and $P$ is fixed. Again with monitoring the following $J^Q_L>0$ and $J^Q_{R,P}<0$  conditions the data for $\eta(P_{max})$ are collected. In this configuration, the two terminal case shows a big value $\simeq 0.9$ for $ZT$  about $\varepsilon=0$ which pertains to the quantum interference effect discussed later.  Likewise, for three-terminal model a great enhancement up to $\simeq 4$ near $\varepsilon=\pm 2$ is observed.
\par
In Fig.~\ref{fig5}, the efficiency at maximum output power normalized over Carnot efficiency $ \eta_{(\textbf{P}_{max})}/\eta_{c}$  is plotted as a function of $ M$ for configuration-$(a)$. Energy is fixed at $ \varepsilon=1.43 eV$. The chemical potentials $ (\delta{\mu}_{L,P})$ are chosen to guarantee maximum output power, i.e., fixing the generalized forces $(X_{L,P}^T)$ to optimize the output power \textbf{P}. For $0\leq M<0.57 $ the system absorbs heat only from reservoir $L$ ($J^Q_L>0,~J^Q_{R,P}<0$); while for $ 0.57\leq M<18.5 $ the system absorbs heat from reservoirs $L$ and $P$  ($J^Q_R<0,~J^Q_{L,P}>0$); and finally, for $18.5\leq M\leq100 $ the system absorbs heat only from reservoir $P$ ($J^Q_P>0,~J^Q_{R,L}<0$). Note that, to follow the linear response regime in our work, we set $\delta T_L=10^{-3}T $ for $ M \leq 2 $ and  $\delta T_P=10^{-3}T$ for $M>2 $. As can be seen from the Fig.~\ref{fig5},  the efficiency at maximum power increases dramatically and then stabilizes,  raising from almost zero at $M=0$ to reach a saturation value $0.12$ about $M=30$. It shows that for the strong and symmetric coupling case ($\gamma_{P} =\gamma_{L}=\gamma_{R} =2.5 eV$), the efficiency at maximum output power for the system under consideration cannot exceed $0.12$, even if high values of $M$ are set.
 \begin{figure}[t]
\includegraphics[width=9cm, height=5cm]{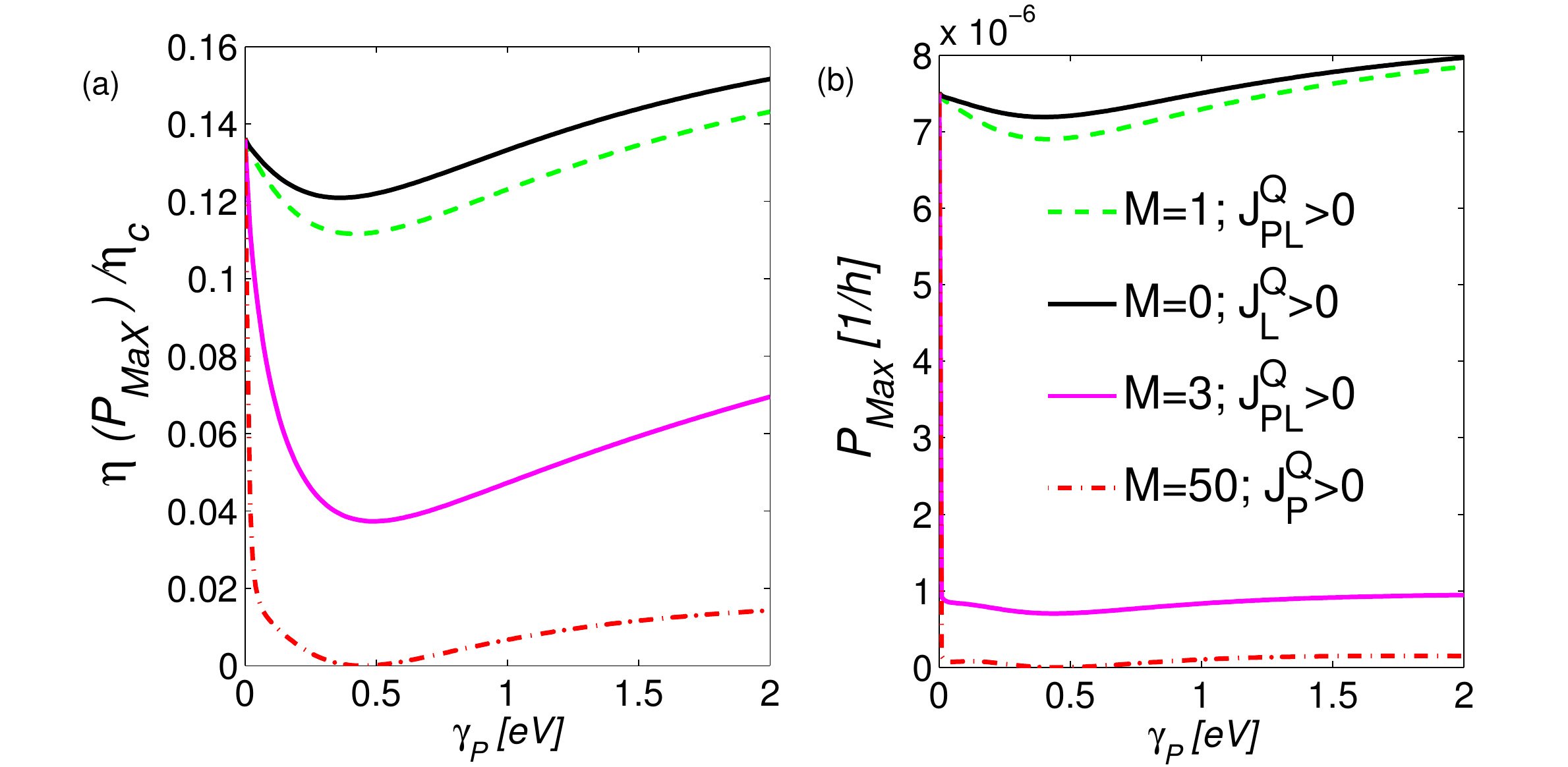}
\caption{(Color online) Efficiency at maximum output power, normalized over the Carnot limit for configuration-$(a)$ as a function of $ \gamma_{P}$ (left panel). Maximum output power extracted by the thermal machine, as a function of $ \gamma_{P}$ (right panel). Different temperature ratio $M=0$ (dotted black line),  $M=1(3)$ (solid green(pink) line), and  $M=50$ (dashed red line) corresponding to $J_L^Q>0 J_{P,R}^Q<0$, $J_R^Q<0 J_{P,R}^Q>0$, and $J_P^Q>0 J_{L,R}^Q<0$ are considered. Other parameters are set as $T=300 K$, $\varepsilon=1.56 eV$, and $\gamma_L=\gamma_R=0.1 eV$.}
\label{fig6}
\end{figure}
 \par
Let's focus on how the efficiency at maximum output power and the maximum output power evolve when the system is driven from a two-terminal to a three-terminal configuration, that is by tuning $\gamma_{P} $. The two terminal system corresponds to $ \gamma_{P}=0 $ and the third terminal is switched on by increasing $ \gamma_{P}$. For simplicity, the coupling strengths of terminals $L$ and $R$ are taken equal to $\gamma$, whereas the coupling strength to terminal $P$ is considered as $\gamma_P$. 
\begin{figure}[t]
\includegraphics[width=9cm, height=5cm]{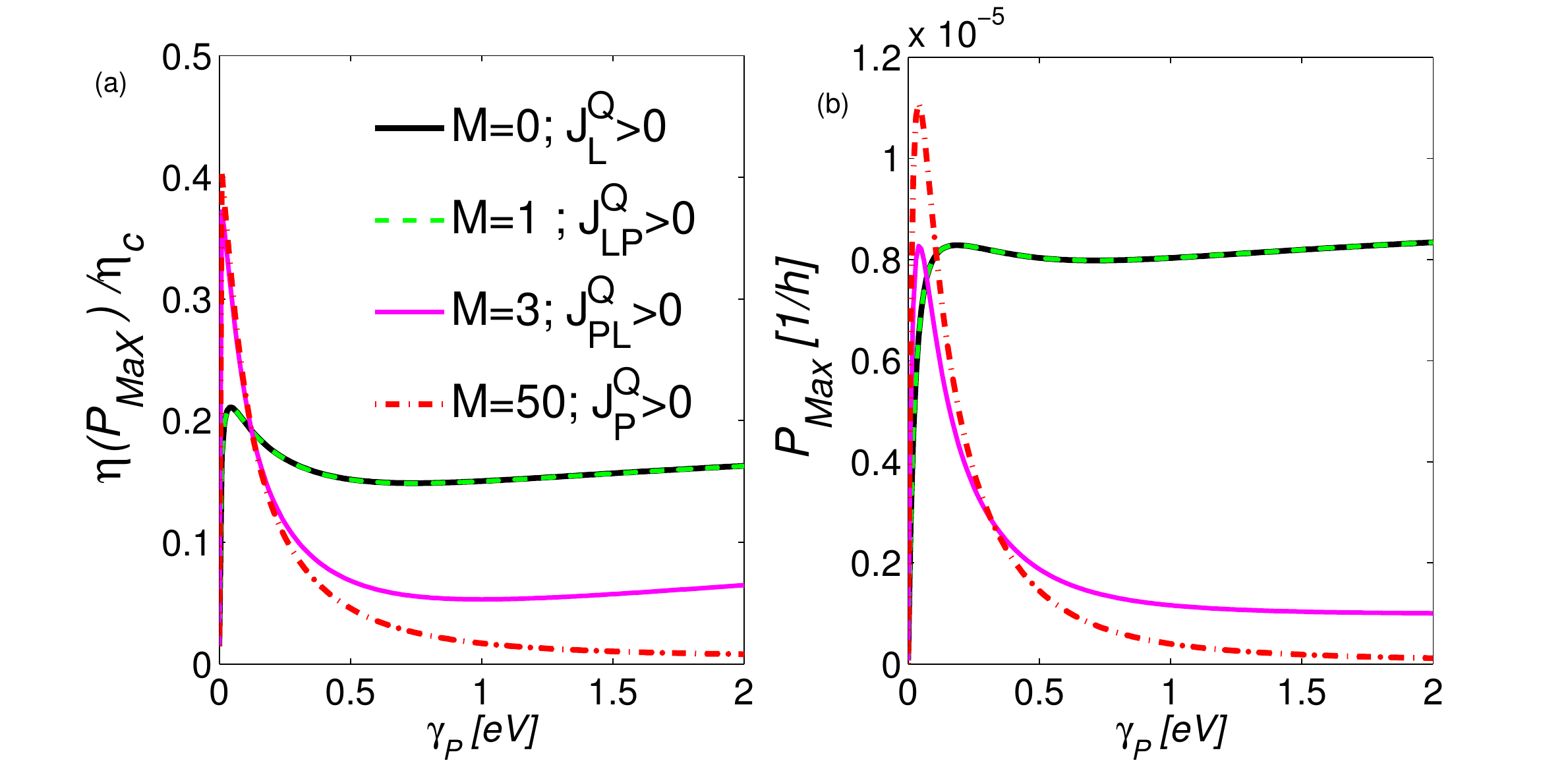}
 \caption{(Color online) Same as Fig.~\ref{fig6}, but for configuration-$(b)$}
\label{fig7}
\end{figure}
\par
Fig.\ref{fig6} drawn for  configuration-$(a)$, we set $M=0$ (dotted-black line), $M=1, (3)$ (green, (pink) line) and $M=50$ (dashed-red line) leading to $\eta_L(\textbf{P}_{max})/\eta_c$ when the system absorbs heat only from contact $L$, $\eta_{LP}(\textbf{P}_{max})/\eta_c$ when the system absorbs heat from contacts $L$ and $P$, and $\eta_P(\textbf{P}_{max})/\eta_c$ when the system absorbs heat only from contact $P$, respectively. It is seen that increasing the coupling $\gamma_P$ strength leads to improvement of the performance. As shown in Fig.~\ref{fig6} both $\eta_L({\bf P}_{max})$ and ${\bf P}_{max}$ increase for all values of $M$ when $\gamma_P\geq0.5 eV$. 
\begin{figure}[t]
 \includegraphics[width=9cm, height=5cm]{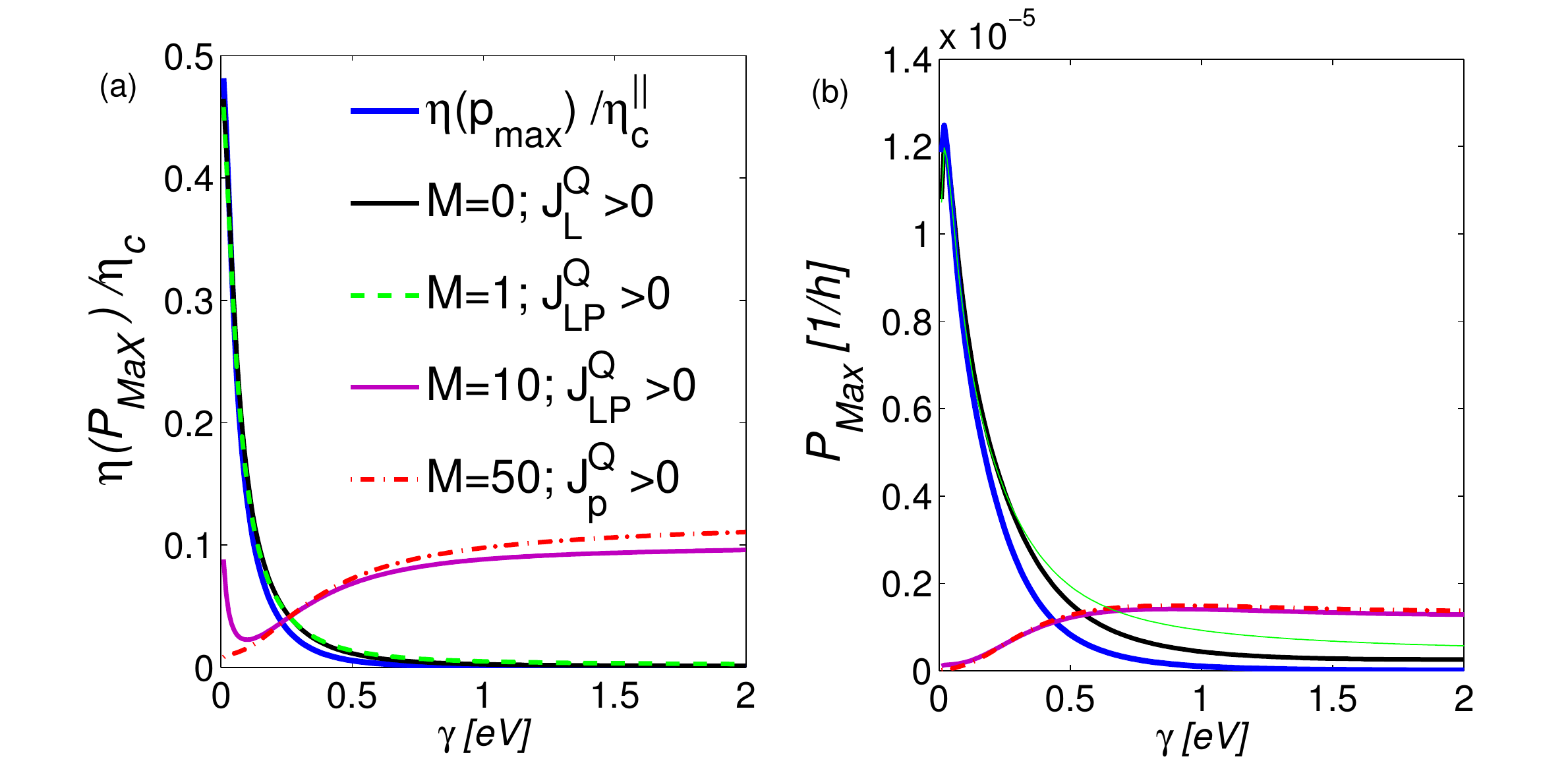}
 \caption{(Color online) $\eta(\textbf{P}_{max})/\eta_{c}$ (left panel)  and $\textbf{P}_{max}$ (right panel) for two-terminal model (blue line) and for configuration-$(a)$ as a function of  $ \gamma_{L}=\gamma_{R}=\gamma $.  $M=0, M=1$ if $\delta T_L=10^{-3}T$ and $M=10, M=50$ if $ \delta T_P=10^{-3}T$.  Other parameters are set as $T=300 K$, $\varepsilon=1.56 eV$, and $\gamma_P=2.0 eV$.} 
\label{fig8}
\end{figure}
\begin{figure}[t]
 \includegraphics[width=9cm, height=5cm]{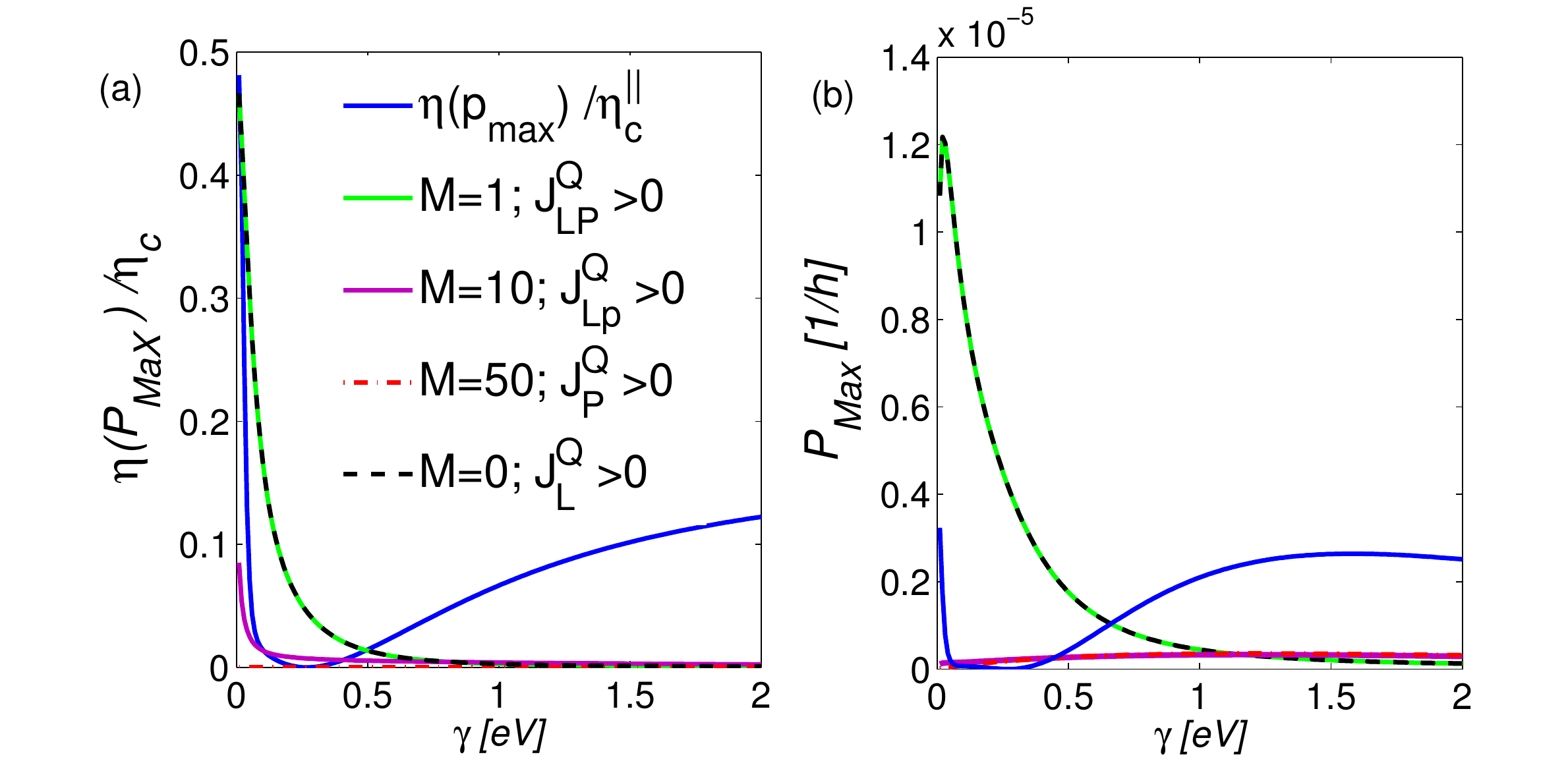}
 \caption{(Color online) Same as Fig.\ref{fig8}, but for configuration-$(b)$} 
\label{fig9}
\end{figure}
\par
In Fig.~\ref{fig7}, we show results for the same quantities but for configuration-$(b)$. In particular, it can be seen that for parameters $M=0$ and $M=1$ the efficiency at maximum output power (left panel) and maximum output power (right panel) increase at small couplings and then fall down steadily, while for $M=50$ and $M=3$ these quantities increase at small couplings $\gamma_P$ and then drop sharply, reaching almost zero at $\gamma_P=2 eV$ for $M=50$.
\par
Next, in Fig.~\ref{fig8} we show results for $ \eta(\textbf{P}_{max}) /\eta_{c}$ (left panel)  and $\textbf{P}_{max}$ (right panel) for the case of two-terminal and configuration-$(a)$ with parameters $M=0 (M=1)$ if $ \delta T_L=10^{-3}T$ and  $M=10 (M=50)$  if $ \delta T_P=10^{-3}T$, as a function of $ \gamma_L=\gamma_R=\gamma$ for two values of $\gamma_P=0$ and $\gamma_P=2.0 eV$.  Note that, for parameters $M=1, (M=0)$ the efficiency coincides with that of a two-terminal system. Results show that the efficiency at maximum output power for $M=0$ and $M=1$ tend to $\eta(\textbf{P}_{max})/\eta_c=0.5$ and $ \eta(\textbf{P}_{max})/\eta_c=0$ in the limit $\gamma\rightarrow0$ and $ \gamma\rightarrow\infty $, respectively, while the trends for $M=10$ and $M=50$ are opposite. For these values of $M$, the efficiency at maximum output power tends to zero in the limit $ \gamma\rightarrow 0$ and reaches  about $0.15$ for strong coupling limit.  The maximum output power saturating the Carnot bound $\eta_{(\textbf{P}_{max})} /\eta_c=0.5$ in the limit $ \gamma\rightarrow0 $  for two-terminal case which is due to delta shaped transmission function resulting in the divergence of the figure of merit $ZT$.The same two contact strong energy dependence of electron transport explains the three-terminal results. Indeed, the three-terminal model is one way to achieve a narrow energy-dependent transport.
In the right panel of  Fig.~\ref{fig8}, the maximum output power show the same trend as efficiency at maximum power for all parameters of $M$. 
\par
In Fig.~\ref{fig9}, we show the same quantities as in Fig.~\ref{fig8}, but for configuration-$(b)$. As can be seen from the figure at small coupling $\gamma$ for $M=1(M=0)$, is greater than the two-terminal setup. In contrast to the two-terminal case which shows a sharp decrease to zero and then a revival feature versus coupling change, the three-terminal model for $M=0, 1$ show an exponential decrease. For other $M$ options, we have not seen eminent trend compared to the two-terminal model, while for configuration-$(a)$ in Fig.~\ref{fig8} at strong coupling regime these $M$ options showed finite value for the efficiency at maximum output power and maximum output power.
\par
Before concluding, we comment on the realization of our funding. A great challenge in building molecule-based electronic devices is making reliable molecular junctions and controlling the electrical current through the junctions. The mechanically controllable break junction technique\cite{a56}  with integrated nanoscale thermocouples\cite{a57}  which is accessible by the current experimental apparatus would one possible way to test the theoretical results in this work.

\section{SUMMARY}\label{sec1}
In summary, with the help of the Landauer-Buttiker formalism and within time-reversal symmetry, the linear response thermoelectric properties through a Benzene molecule coupled to three non-magnetic electrodes is studied. We consider the two well-known para and meta configurations and added a third terminal in the ortho position. We focus on the temperature difference and coupling strength to find a setup which drives the system to a condition which can show better efficiency than the conventional two-terminal junction. By introducing temperature difference parameter $M$ between terminals and tuning its value, we found cases where the efficiency at maximum output power and maximum output power of three-terminal configuration exceed the typical two-terminal model. Particularly, under $M=0.1$ the figure of merit is dramatically influenced by the temperature for both configurations considered.
\appendix
\section{Linear response and Onsager reciprocal relations}\label{apeA}
The four well-known Onsager matrix $\bm{{\cal L}}$ elements, the electrical conductance $G$, the Peltier coefficient$ \amalg $, the thermal conductance $K$ and the Seebeck coefficient $S$ under certain limitations gauge the transport properties of the system
The case for multi-terminal setup is extend to the introduction of nonlocal coefficients, which describe the infeluence of bias driven between two terminals on the another terminal. By assuming linear regime, namely small thermodynamic forces, the relationship between fluxes and forces can be as $(J=\bm{{\cal L}}X)$
\begin{equation}
\begin{bmatrix}
J_L^N\\
J_L^Q\\
J_P^N\\
J_P^Q
\end{bmatrix}
=
\begin{bmatrix}
{\cal L}_{11} &{\cal L}_{12} & {\cal L}_{13} & {\cal L}_{14}\\
{\cal L}_{21} & {\cal L}_{22} & {\cal L}_{23} & {\cal L}_{24}\\
{\cal L}_{31} & {\cal L}_{32} & {\cal L}_{33} & {\cal L}_{34}\\
{\cal L}_{41} & {\cal L}_{42} & {\cal L}_{43} & {\cal L}_{44}
\end{bmatrix}
\begin{bmatrix}
X_L^{\mu}\\
X_L^T\\
X_P^{\mu}\\
X_P^T
\end{bmatrix}
\label{eqA1}
\end{equation}
with time-reversal invariance of the equations of motion, Onsager found fundamental relations, known as Onsager reciprocal relations for the cross coefficients of the Onsager matrix ${\cal L}_{ij}={\cal L}_{ji}$. Due to the positivity of the entropy production rate, such a matrix has to be semi-positive definite (i.e. $\bm{{\cal L}}\ge0$) and that it can be used to introduce a two-terminal configuration that connect electrode $ L $ with electrode $ R $ by setting $L_{j3}$ = $L_{j4}$ =$ L_{3j}$ = $L_{4j}$ = $0$  for $j=1,2,3,4$.
\par
For a two-terminal system,  the Seebeck coefficient $ S $ relates the voltage difference $\delta V$  between the terminals to their temperature difference $ \delta T$ with an open circuit condition ( zero charge current), the electrical conductance $(G)$ describes  the electric current dependence on the voltage differences between the two-terminal when two terminals have the same temperature, the thermal conductance $(K)$ describes the heat current dependence on the temperature difference $ \delta T $ under the assumption that no net charge current is flowing through the system, and the Peltier effect $ (\Pi) $ relates the heat current to the charge current under isothermal condition. An extension to the multi-terminal scenario is achieved by introducing the matrices of elements
\begin{eqnarray}
G_{ij}&=&\Bigg[\frac{e^2J_i^N}{\delta\mu_j}\Bigg]_{\delta T_k=0~\forall k};\quad  K_{ij}=\Bigg[\frac{J_i^Q}{\delta T_j}\Bigg]_{J^N_k=0~\forall k}\nonumber\\
 S_{ij}&=&-\Bigg[\frac{\delta\mu_i}{e\delta T_j}\Bigg]_{J_k=0~\forall k};\quad \Pi_{ij}=\Bigg[\frac{J_i^Q}{eJ^N_j}\Bigg]_{\delta T_k=0~\forall k}\nonumber\\
\label{eqA2}
\end{eqnarray}
with local ($i=j$) and non-local ($i\ne j$) coefficients, $e$ being the electron charge.
\\
\section{Scattering approach in linear response regime: The Onsager coefficients}\label{apeC}
The heat and particles currents through a non-interacting conductor can be described by the multi-terminal Landauer-Buttiker approach\cite{a53}. Assuming that all non-coherent processes, phase breaking and dissipative, happen in the terminals, the charge and heat currents from terminal $L$ (reservoir) are given by \cite{a54,a55}
\begin{eqnarray}
J_L^N&=&\frac{1}{h}\int dE\sum_{j\ne L}\left[{\cal T}_{jL}(E)f_L(E)-{\cal T}_{Lj}(E)f_j(E)\right]\nonumber\\
J_L^Q&=&\frac{1}{h}\int dE(E-\mu_L)\sum_{j\ne L}\left[{\cal T}_{jL}(E)f_L(E)-{\cal T}_{Lj}(E)f_j(E)\right]\nonumber\\
\label{eqC1}
\end{eqnarray}
where $f_j(E)$=$\left[\exp\left[(E-\mu_j)/k_BT_j\right]+1\right]^{-1}$ is the Fermi function and ${\cal T}_{ij}$ is the transmission probability from terminal $L$ to terminal $j$. Analogous expressions can be defined for $J_P^N$ and $J_P^Q$, provided the terminal $L$ is substituted by $P$. 
\par
For a three-terminal configuration, one choose the electrode $R$ as the reference $(\mu_{R}$=$\mu$=$0, T_{R}$=$T)$. As in the previous sections, one set $ \mu_{L,P}$=$ \mu+\delta{\mu}_{L,P}$,  $ T_{L,P}$=$ T+\delta{T}_{L,P}$. The Onsager coefficients ${{\cal L}_{ij}}$ can be obtained from the linear response expansion ($J=\bm{{\cal L}}X$) of the currents $J_i^N$ and $J_i^Q (i=L,P)$ ;
\begin{eqnarray}
{\cal L}_{11}&=&-\frac{T}{h}\int_{-\infty}^{\infty}\Big[f'(E)\sum_{\gamma\ne L} {\cal T}_{L\gamma}(E)\Big]~dE\nonumber\\
{\cal L}_{12}&=&-\frac{T}{h}\int_{-\infty}^{\infty}\Big[f'(E)(E-\mu)\sum_{\gamma\ne L} {\cal T}_{L\gamma}(E)\Big]~dE={\cal L}_{21}\nonumber\\
{\cal L}_{22}&=&-\frac{T}{h}\int_{-\infty}^{\infty}\Big[f'(E)(E-\mu)^2\sum_{\gamma\ne L} {\cal T}_{L\gamma}(E)\Big]~dE\nonumber\\
{\cal L}_{13}&=&\frac{T}{h}\int_{-\infty}^{\infty}\Big[f'(E) {\cal T}_{LP}(E)\Big]~dE={\cal L}_{31}\nonumber\\
{\cal L}_{14}&=&\frac{T}{h}\int_{-\infty}^{\infty}\Big[f'(E)(E-\mu){\cal T}_{LP}(E)\Big]~dE={\cal L}_{41}\nonumber\\
{\cal L}_{24}&=&\frac{T}{h}\int_{-\infty}^{\infty}\Big[f'(E)(E-\mu)^2{\cal T}_{LP}(E)\Big]~dE={\cal L}_{42}\nonumber\\
{\cal L}_{23}&=&\frac{T}{h}\int_{-\infty}^{\infty}\Big[f'(E)(E-\mu){\cal T}_{LP}(E)\Big]~dE={\cal L}_{32}\nonumber\\
{\cal L}_{33}&=&-\frac{T}{h}\int_{-\infty}^{\infty}\Big[f'(E)\sum_{\gamma\ne P} {\cal T}_{P\gamma}(E)\Big]~dE\nonumber\\
{\cal L}_{34}&=&-\frac{T}{h}\int_{-\infty}^{\infty}\Big[f'(E)(E-\mu)\sum_{\gamma\ne P} {\cal T}_{P\gamma}(E)\Big]~dE={\cal L}_{43}\nonumber\\
{\cal L}_{44}&=&-\frac{T}{h}\int_{-\infty}^{\infty}\Big[f'(E)(E-\mu)^2\sum_{\gamma\ne P} {\cal T}_{P\gamma}(E)\Big]~dE
\label{eqC3}
\end{eqnarray}
where $f'(E)$ is the Fermi-Dirac distribution derivative with respect to the energy and $ T $ is the temperature. As mentioned before, within the time reversal symmetry $ {{\cal L}_{ij}}$=${{\cal L}_{ji}}$.
\end{document}